# Are LLM Agents the New RPA? A Comparative Study with RPA Across Enterprise Workflows

Petr Průcha[1][0000-0003-2197-7825] Michaela Matoušková[1][0009-0008-3451-5260] and Jan Strnad[2]

[1] Technical University of Liberec, Liberec, Czechia
`petr.prucha@tul.cz, michaela.matouskova@tul.cz`
[2] Pointee Inc., Delaware, USA

**Abstract.** The emergence of large language models (LLMs) has introduced a new paradigm in automation: LLM agents or Agentic Automation with Computer Use (AACU). Unlike traditional Robotic Process Automation (RPA), which relies on rule-based workflows and scripting, AACU enables intelligent agents to perform tasks through natural language instructions and autonomous interaction with user interfaces. This study investigates whether AACU can serve as a viable alternative to RPA in enterprise workflow automation. We conducted controlled experiments across three standard RPA challenges data entry, monitoring, and document extraction comparing RPA (via UiPath) and AACU (via Anthropic's Computer Use Agent) in terms of speed, reliability, and development effort. Results indicate that RPA outperforms AACU in execution speed and reliability, particularly in repetitive, stable environments. However, AACU significantly reduces development time and adapts more flexibly to dynamic interfaces. While current AACU implementations are not yet production-ready, their promise in rapid prototyping and lightweight automation is evident. Future research should explore multi-agent orchestration, hybrid RPA-AACU architectures, and more robust evaluation across industries and platforms.

**Keywords:** Robotic Process Automation, RPA, LLM agent, Computer use, AI Automation, Agentic Automation

## 1 Introduction

Currently, large language models (LLMs) and generative AI are rapidly gaining attention in the field of automation. Recent advances in LLMs have enabled a new paradigm known as *AI agentic automation*. In this approach, intelligent agents can perform specific tasks by interacting with digital systems in a way that resembles how a human user would operate a computer. A specific subset of this, referred to here as *AI Agentic Automation with Computer Use* (AACU) or LLM agents with computer use, allows agents to use software tools, interfaces, and APIs autonomously based on high-level natural language instructions.

AACU shares several traits with traditional Robotic Process Automation (RPA). Both technologies aim to automate repetitive digital tasks to reduce human effort, minimize errors, and improve operational efficiency. However, they differ fundamentally in implementation. RPA requires well-defined workflows and low-code or coded instructions to automate specific tasks. In contrast, AACU uses natural language prompts as input. These prompts are dynamically interpreted and translated into executable actions by the agent, reducing the need for explicit programming and enabling faster, more adaptive automation.

The literature suggests that AACU may eventually replace RPA in certain use cases [1–3]. While RPA has demonstrated clear benefits such as cost reduction, improved process quality, and error mitigation across various industries it still relies heavily on structured rules and manual configuration [4–6]. AACU, by contrast, represents a more flexible and scalable solution. It promises faster deployment and greater adaptability by leveraging the reasoning and language understanding capabilities of LLMs [7]. In this sense, AACU could potentially accelerate process automation by minimizing the technical barrier traditionally associated with automation tools.

Despite the growing excitement around LLM-based agents, there is a noticeable gap in comparative research between AACU and RPA. Current discourse is largely driven by hype rather than empirical evidence. It remains unclear whether agentic automation can reliably perform the same tasks as RPA, or even surpass it in terms of performance, scalability, and usability. This research aims to investigate the current state of AACU and compare it with the established capabilities of RPA. The goal is to evaluate whether intelligent agents can serve as a viable replacement for RPA in business process automation and to assess their strengths, limitations, and practical implications.

## 2    Related work

The term *"agentic automation"* does not currently yield any relevant literature focused specifically on business process automation in academic databases such as Scopus or Web of Science. However, emerging research has begun to explore the integration of AI agents into enterprise processes using new technologies such as Retrieval-Augmented Generation (RAG) and Model Context Protocols (MCP) [8, 9].

Recent research suggests a shift from traditional RPA and intelligent automation toward more advanced forms of automation that operate directly from human language inputs (Chakraborti et al., 2022; Rizk et al., 2020; Gotthardt et al., 2020). Intelligent agents are increasingly seen as the future of robotic process automation, largely due to advances in artificial intelligence (AI) and machine learning (Siderska et al., 2023; Huo et al., 2023; Rizk et al., 2022; Afrin et al., 2025). Although the trend is moving toward greater agent autonomy, some systems still rely on RPA to execute specific tasks. This continued use of RPA is due to its high efficiency in interacting with computer environments and its ability to simulate user actions reliably. Furthermore, when process errors occur, RPA can generate understandable code that helps users identify and correct issues more easily (Afrin et al., 2025).

During the literature review, several resources related to agentic automation were found on preprint servers. Given the novelty of this topic, it is understandable that much of the research has not yet undergone formal peer review or been published in established academic journals. One such study by Gaurav Samdani et al. focuses on improving organizational processes through the use of agentic automation, demonstrating early evidence of its potential benefits in real-world business environments [10].

Sapkota et al. provide a taxonomy and clear description of AI agents and agentic AI, illustrating the differences between these two concepts with practical examples [11]. Similarly, Ye et al. (2023) introduce a new paradigm for agentic process automation through the development of the PROAGENT tool, which can automate processes based on user instructions [12]. This tool demonstrates the potential of AI agents and their ability to interact with computer systems to carry out complex tasks autonomously.

Technology companies are currently racing to be the first to offer robust Agentic Automation with Computer Use (AACU) platforms that can complete tasks and processes based solely on human instructions. However, many of these tools remain in closed beta testing or are not yet publicly accessible worldwide. For example, some of the most advanced AACU tools, such as those developed by OpenAI, are currently only available in the United States. Microsoft has also announced a computer-use version of Copilot, but it has not yet been released to the public. Fortunately, in the past year, Anthropic released a publicly available computer-use tool via its open repositories. According to Lamanna, these tools are designed to perform actions through user interfaces, addressing many of the same tasks traditionally handled by RPA solutions [13].

Given the growing interest, media buzz, and rapid technological developments, we propose a set of hypotheses to evaluate the capabilities and practical viability of AACU in comparison to traditional RPA systems.

*H1: The AACU performs automation faster than RPA.*
*H2: The AACU is more reliable than RPA.*
*H3: The AACU requires less development time than RPA.*

## 3  Process Overview

For this study, we selected three processes, described in detail below. These processes are representative of typical RPA practices and are sometimes used in RPA competitions. All three were sourced from the well-known website https://rpachallenge.com, which features five example processes designed to simulate various operations typically performed within organizations and now commonly automated using RPA.

**Process 1 – The RPA Challenge**
Process 1 (P1) is widely recognized in the RPA community as "The RPA Challenge." To successfully complete this challenge, an RPA or AACU must extract data from a spreadsheet and input it into specific fields within a web form. The difficulty lies in the fact that the form layout changes with each round, making the automation more complex. This process is illustrated in Figure 1.

Although this type of dynamic layout change is uncommon in real-world enterprise systems, the challenge represents a typical RPA use case: transferring data from Excel into other systems. The process requires the RPA or AACU to accurately read and map employee data from an Excel file to the corresponding fields, even as the form structure shifts dynamically.

**Process 2 – RPA Challenge Stock Market**
Process 2 (P2) is a lesser-known challenge from the same domain as the RPA Challenge (P1), available at https://rpachallenge.com/assets/rpaStockMarket/index.html. Unlike P1, this challenge does not include formal instructions. After researching how others approached it—primarily through demonstration videos—we tailored the challenge to test web scraping or screen scraping capabilities, along with simple logic implementation.

In our version of the task, the RPA bot or AACU must navigate to the challenge website and select the company "Exenon UI Pharma" from a dropdown menu. Once selected, the AACU must monitor the fictional stock price displayed on the page. If the price drops below $70, it should notify the user that this threshold has been crossed.

This process evaluates the AACU's ability to extract the correct data from a website, continuously monitor changes, and trigger a user-facing message when a predefined condition is met. It is important to note that the stock prices on the site are static—they remain the same each time the page is loaded, ensuring consistency across test runs.
The process is illustrated in Figure 2

**Process 3 – RPA challenge Invoices – Simplified for AACU**
Process 3 (P3) is another lesser-known challenge, accessible through navigation on the RPA Challenge website. The specific link to this task is https://rpachallengeocr.azurewebsites.net. While this challenge includes clear instructions, we tailored it to better align with the capabilities of the agent (i.e., the computer-based tool used in our study). The rationale for this adaptation is discussed in greater detail in the Methodology and Discussion sections. Additionally, it's worth noting that the Invoice Challenge is maynot fully functional or actively maintained, as reflected in the current state of its associated [14].

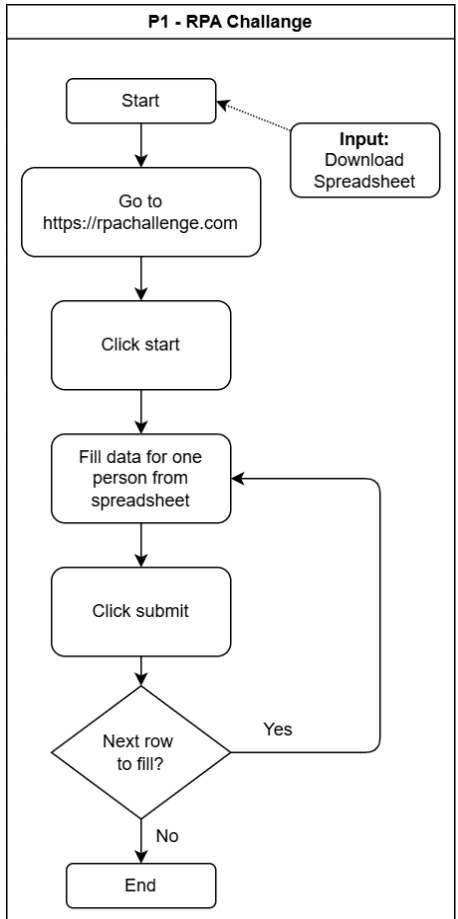
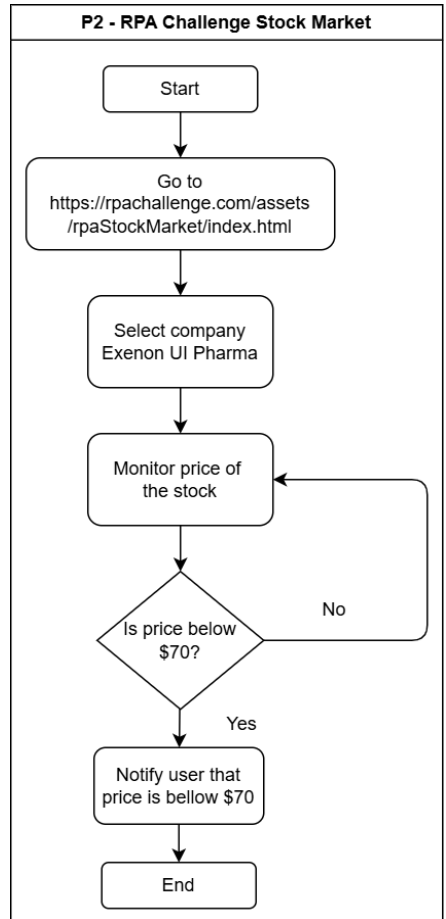

**Fig. 1.** Visualisation of process 1 the RPA challange

**Fig. 2.** Visualisation of process 2 the RPA challange stock market

The visualization of the Invoice Challenge is shown in Figure 3. The process can be described as follows:

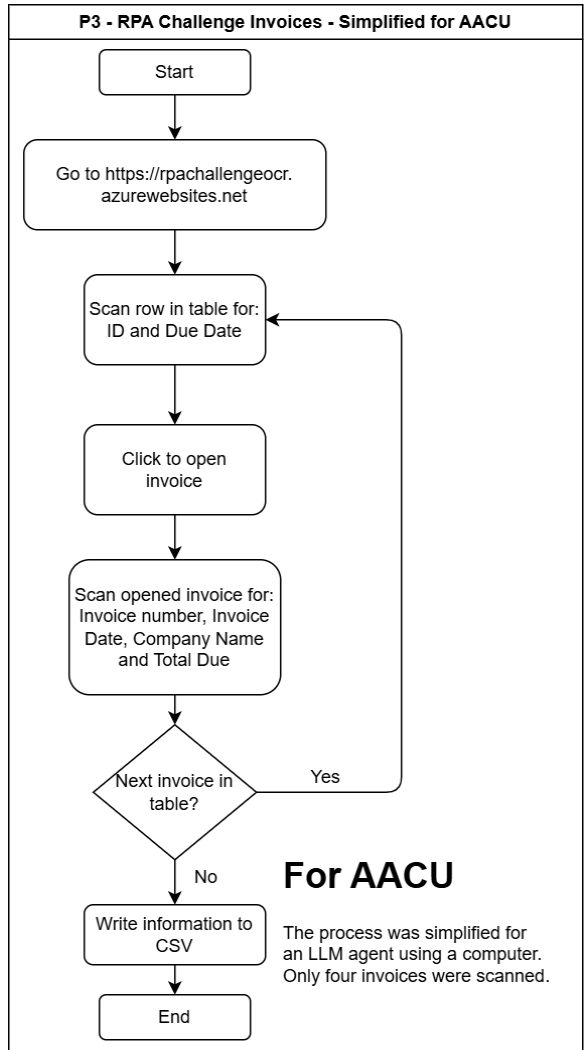

**Fig. 3.** Visualisation of process 3 the RPA invoice challange

The RPA bot or AACU navigates to https://rpachallengeocr.azurewebsites.net, where it interacts with a table of invoice records. From this table, the agent must extract the Invoice ID and Due Date for each entry. Additionally, each row contains a link to view the invoice as an image. From the image, the agent must further extract the Invoice Number, Invoice Date, Company Name, and Total Due.

There are two fictional companies featured in this challenge, each using a slightly different invoice layout (see Figure 3). Upon each reload of the page, the order of invoices in the table is randomized, meaning it's unlikely to encounter the same order twice.

Therefore, the agent must correctly identify each invoice and extract the required information from the corresponding image.

The table typically displays four invoices per page, with additional invoices accessible on subsequent pages. However, due to tool limitations, we limited the scope of this challenge to the first page only, which contains four invoices.

The extracted data for each invoice was written to a .csv file, following the output format described in the challenge instructions. A sample output is illustrated in Figure 4.

```
ID,DueDate,InvoiceNo,InvoiceDate,CompanyName,TotalDue
5jef1y8yx4t8yupbpo3fzg,25-02-2019,10021,13-02-2019,Sit Amet Corp.,1234.40
g11rzk7loegnitbclwnh8k,06-04-2019,223543,30-01-2019,Aenean LLC,976.00
```

**Fig. 4.** Example of output from invoice challenge P3

## 4 Methodology

To test the hypotheses outlined in Chapter 2: Related Work, we conducted experiments using three processes described in detail in Chapter 3. For the RPA-based automation, we used UiPath Studio 2023.4.0 Community Edition. To represent agentic automation, we utilized Anthropic's Computer Use Agent, a tool capable of interacting with the graphical user interface (GUI) to perform tasks.

The Anthropic Computer Use Agent is currently in public beta and not recommended for production environments due to safety limitations. In accordance with Anthropic's guidelines, we followed the official setup instructions. The agent ran in a controlled Docker-based environment provided by Anthropic, using Ubuntu 22.04.5 LTS as the base operating system. The environment included default desktop applications such as Firefox, LibreOffice, and Gedit.

The specific version of the computer-use agent was:
- GitHub commit: 99502f5
- Model: claude-sonnet-4-20250514
- API Provider: Anthropic (default settings)
- Tool Version: computer_use_20250124
- Max Output Tokens: 16,384
- Image Input: Enabled (sends 3 most recent screenshots per action)

Each process (P1–P3) was tested using both UiPath and the agentic automation tool. Each scenario was executed 10 times to assess the consistency and reliability of the technology — with the exception of Process 1 (P1), where the agent was only run once. In the case of P1, the Anthropic agent was unable to complete the full process due to current tool limitations. After several actions, the application would freeze, fail to accept further input, or attempt to reconnect unsuccessfully. Refreshing the session caused the agent to lose context and require re-prompting from the beginning.

As mentioned earlier, Process 3 (P3) was shortened to fit within the current context limitations of the agent. Even in this reduced form, the process remained challenging. The agent typically began by opening Firefox to the starting URL and preparing a structured note using Gedit, but maintaining context throughout the entire workflow proved difficult.

Initially, simple natural language prompts were used to instruct the agent. However, due to frequent application freezes, we switched to providing complete, step-by-step process instructions in a single prompt. These prompts remained relatively simple, describing the sequence of actions required. Below is an example of the prompt used for Process 2 (P2):

*1)Open Firefox*
*2)Go to https://rpachallenge.com/assets/rpaStockMarket/index.html*
*3)Click Select*
*4)Select the Exenon Ui Pharma*
*5)Tell me when the price drop below 70*

For Process 3 (P3), we initially applied the same prompting method used in previous scenarios. However, the original prompt proved to be less effective. To improve performance, we asked Claude Sonnet 4 to rewrite and optimize the prompt. The revised version led to better execution. Due to its length, the full prompt is provided in Appendix 1 and is also available in the public repositories associated with this research.
In addition to evaluating process success, we also measured development time defined as the time taken until the first successful run of each automation. This includes both:
- RPA development time using UiPath, and
- Prompt engineering time for the AACU.

Timing was recorded manually using a stopwatch, so there may be some measurement inaccuracies. Each process was developed once per technology, and thus the reported times reflect a single measurement per scenario.
In the case of RPA, development time was tracked using UiPath's built-in tools, while for the agent, a stopwatch was used to measure prompting time. All performance data–including RPA development metrics and screenshots from agent executions, are available in the corresponding GitHub repositories[1].
It is worth noting that some screenshots were not captured due to application instability during agent runs. In cases where the interface froze or misbehaved, the agent could not proceed, and screenshots could not be taken. This behavior is documented in the public repository.
To validate the research hypotheses, the following statistical methods were applied:
- For Hypothesis 1, we used a Welch's t-test (two-sample t-test assuming unequal variances).
- For Hypothesis 2, we applied Fisher's Exact Test, as the nature of the data did not meet the assumptions required for a Chi-Square test.
- For Hypothesis 3, due to limitations in measurement reliability, we chose to interpret the results descriptively without applying statistical significance testing.

---

[1]Github repository: https://github.com/Scherifow/LLM-agents-Research.git

## 5  Results

The results of the experimental measurements for each process are summarized in Table 1, which presents the average execution time (in seconds), the number of successful and unsuccessful runs, and the performance of both RPA and the Agent. Each process was executed 10 times per technology, except for P1 with the AACU, where only one run was performed due to technical limitations.

**Table 1.** Results of duration and success rate of the challanges per technology

| Process | RPA time (s) | Successful X Unsuccesful run RPA | AACU time (s) | Successful X Unsuccesful run AACU |
|---|---|---|---|---|
| P1 | 139.8 | 10 / 0 | X | 0 / 1 |
| P2 | 53.9 | 10 / 0 | 109.8 | 9 / 1 |
| P3 | 20 | 10 / 0 | 202.8 | 6 / 4 |

*H1: The AACU performs automation faster than RPA.*
To evaluate this hypothesis, we conducted a Welch's t-test (two-sample t-test assuming unequal variances) comparing the execution times of RPA and the AACU for P2 and P3. The results are presented in Table 2.

**Table 2.** Results of welch t-test for hypothesis 1

| Process | t-test | p-value |
|---|---|---|
| P2 | -3.3841 | 0.0096 |
| P3 | -17.30 | 0.0000118 |

The first hypothesis (H1), which states that the Agent automates processes faster than RPA, is not accepted based on the t-test results. For process P2, the p-value of 0.0096 is less than the significance level $\alpha = 0.05$, and the absolute t-statistic (3.384) exceeds the critical t-value (2.306), leading to the rejection of the null hypothesis. Similarly, for process P3, the p-value (0.0000118) is well below 0.05, and the t-statistic (17.30) is greater than the critical value (2.571), further confirming rejection. These results indicate that RPA performs faster than the AACU, thus disproving the hypothesis that the AACU is faster in automating processes.

*H2: The AACU is more reliable than RPA.*

To evaluate reliability (measured as success rate), we applied Fisher's Exact Test, which is appropriate given the small sample size and categorical nature of the data. The results are displayed in Table 3.

Table 3. Results Fisher's Exact test of hypothesis 2

| Process | Fisher's Exact Test Odds Ratio | p-value |
|---------|-------------------------------|---------|
| P2      | 0                             | 1       |
| P3      | 0                             | 0.0866  |

While the odds ratios suggest that the AACU had lower reliability (0 indicates no successful outcomes relative to RPA), the p-values indicate that these differences are not statistically significant at $\alpha = 0.05$. Thus, we cannot confirm that the AACU is either better or worse than RPA in terms of reliability based on the current data.

**Development Time Analysis**

We tracked the total time required to develop each process using both RPA and the AACU. This measurement includes scripting, prompt creation, debugging, and achieving the first successful run. The development times are approximate and summarized in Table 4.

All times are rounded to the nearest minute. Our results show that development with RPA generally took more time. For Process 1 in RPA, delays occurred due to incorrectly selected web elements that needed to be reselected, which took approximately 15 minutes. In Process 2, issues arose from a dropdown menu that required fixing to enable a successful run, adding about 10 minutes to development. For Process 3, additional time was needed because the developer had to learn how to use OCR technology effectively and parse the resulting data. Moreover, an extra hour was spent converting the data into the correct format due to a developer error. Completing the full code challenge via video tutorials took roughly one hour in total [15].

Table 4. Duration development of successful automation

| Process | RPA development time (min) | AACU development (prompting) time (min) |
|---------|----------------------------|------------------------------------------|
| P1      | ≈ 40                       | X                                        |
| P2      | ≈ 38                       | ≈ 10                                     |
| P3      | ≈ 240                      | ≈ 15                                     |

RPA delays are summarized as follows:
- P1: ~15 minutes correcting element selectors
- P2: ~10 minutes fixing dropdown menu handling
- P3: ~1 hour studying documentation + 1 hour correcting OCR output formatting

*H3: The AACU requires less development time than RPA.*
Due to the single-sample measurement for each process, no statistical test could be applied to validate Hypothesis 3. However, based on observed results in Table 4, it appears that the AACU required significantly less development time for P2 and P3. Despite this encouraging result, further research with multiple samples and controlled trials is needed to confirm this hypothesis.

## 6   Discussion and Limitations

The results clearly favor RPA. However, it is important to note that RPA has benefited from over a decade of development, during which many challenges and issues have been mitigated [16, 17]. The AACU demonstrated strong capabilities in screen element recognition and OCR. For example, during the failed execution of P1 (the RPA challenge), changes in the form layout did not pose a problem for the LLM agent. It was precise and consistently able to locate the appropriate fields for data input. Similarly, the LLM performed reliably when processing invoices, accurately scanning tables and other structured information. The agent also consistently clicked in the correct locations, indicating a high level of precision in interface interaction.

The main issue, aside from the overall process duration, was the occasional unpredictability of the ACCU. For instance, it sometimes launched LibreOffice unnecessarily, particularly in P3, which typically led to failed process execution. Unfortunately, we currently lack an easy way to configure the model's "temperature" setting to ensure deterministic behavior. While the process was generally repeatable, ensuring reliability and diagnosing errors remains a challenge. This reflects a broader issue with generative AI: it operates as a black box, is prone to hallucinations (albeit improving), and makes error recognition difficult [18–20]. To ensure successful automation, additional verification steps may be necessary to maintain quality and consistency[21, 22].

Another important aspect to consider is the cost of AACU. A single execution of P3 cost approximately $0.28. When accounting for the development time and salaries of RPA developers, AACU could become a compelling option–especially for smaller automations that do not run frequently. This may lead to a reconsideration of van der Aalst's framework for selecting automation technologies see Figure 5 [23]. Agentic automation thus expands the automation toolbox beyond APIs, workflows, and RPA, as discussed by [24].

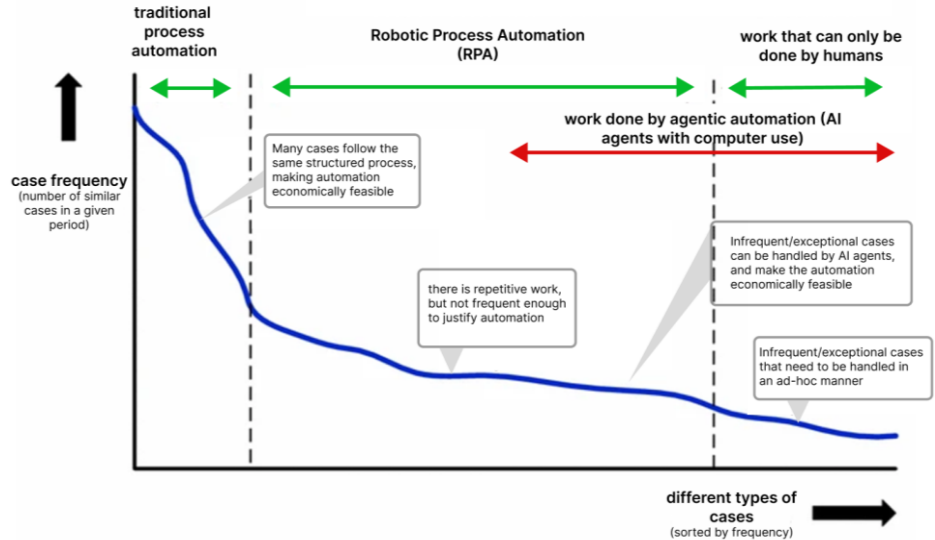

**Fig. 5.** Visualization of [23] feasibility of RPA with incorporation of AACU and AI agents.

**Limitation of the research**
Every research study has its limitations, and this one is no exception. One limitation acknowledged by the authors is the use of only a single RPA technology and a single agentic computer-use technology. Unfortunately, due to current availability constraints, no alternative agentic automation tools were accessible for broader testing. Many of these technologies are based in the United States or are not yet open to a wider audience. This limitation also presents an opportunity for future research to explore and compare multiple agentic automation platforms.

Another limitation lies in the measurement of RPA development time, which was based on a single developer with a specific level of experience. While mistakes made during development (as discussed in Chapter 5) were deducted to ensure a fairer comparison, the overall development time closely resembled that of publicly available video tutorials for the same processes. Nevertheless, the inclusion of these mistakes also reflects the reality of developing automation in unfamiliar environments—where errors and subsequent corrections are an expected part of the process. Importantly, these factors do not undermine Hypothesis 3, which states that prompting an LLM agent is significantly faster than developing an RPA automation.

A final, minor limitation stems from the use of a stopwatch to track time. While this method lacks absolute precision, the authors made every effort to ensure accurate measurements. This limitation is not expected to impact the overall results or conclusions of the study.

## 7 Conclusion

This study offers one of the first empirical comparisons between agentic automation with computer use (AACU) and traditional RPA across standardized workflow tasks. While RPA outperformed AACU in both speed and reliability, AACU demonstrated notable strengths in flexibility and significantly reduced development time. These findings suggest that AACU, though not yet ready to replace RPA in mission-critical workflows, holds substantial promise—particularly in contexts where rapid deployment and adaptability outweigh execution speed. Future research should explore broader process categories, alternative agentic tools, and hybrid architectures that combine the strengths of both technologies.

**Acknowledgment:** This research was made possible thanks to the Technical University of Liberec and the SGS grant number: SGS-2025-1563. This research was conducted with the help of Pointee.

**Disclosure of Interests.** The authors have no competing interests to declare that are relevant to the content of this article.

## References

1. Chakraborti, T., Rizk, Y., Isahagian, V., Aksar, B., Fuggitti, F.: From Natural Language to Workflows: Towards Emergent Intelligence in Robotic Process Automation. In: Marrella, A., Matulevicius, R., Gabryelczyk, R., Axmann, B., Vuksic, V.B., Gaaloul, W., Stemberger, M.I., Ko, A., and Lu, Q. (eds.) Business Process Management: Blockchain, Robotic Process Automation, and Central and Eastern Europe Forum. pp. 123–137. Springer International Publishing Ag, Cham (2022). https://doi.org/10.1007/978-3-031-16168-1_8.
2. Jansen, J.A., Manukyan, A., Al Khoury, N., Akalin, A.: Leveraging large language models for data analysis automation. PLoS ONE. 20, e0317084 (2025). https://doi.org/10.1371/journal.pone.0317084.
3. Agostinelli, S., Marrella, A., Mecella, M.: Towards Intelligent Robotic Process Automation for BPMers, https://arxiv.org/abs/2001.00804, (2020). https://doi.org/10.48550/ARXIV.2001.00804.
4. Wewerka, J., Reichert, M.: Robotic process automation - a systematic mapping study and classification framework. Enterprise Information Systems. 17, 1986862 (2023). https://doi.org/10.1080/17517575.2021.1986862.
5. Enriquez, J.G., Jimenez-Ramirez, A., Dominguez-Mayo, F.J., Garcia-Garcia, J.A.: Robotic Process Automation: A Scientific and Industrial Systematic Mapping Study. IEEE Access. 8, 39113–39129 (2020). https://doi.org/10.1109/ACCESS.2020.2974934.
6. Syed, R., Suriadi, S., Adams, M., Bandara, W., Leemans, S.J.J., Ouyang, C., ter Hofstede, A.H.M., van de Weerd, I., Wynn, M.T., Reijers, H.A.: Robotic Process

Automation: Contemporary themes and challenges. Computers in Industry. 115, 103162 (2020). https://doi.org/10.1016/j.compind.2019.103162.
7. Hughes, L., Dwivedi, Y.K., Malik, T., Shawosh, M., Albashrawi, M.A., Jeon, I., Dutot, V., Appanderanda, M., Crick, T., De', R., Fenwick, M., Gunaratnege, S.M., Jurcys, P., Kar, A.K., Kshetri, N., Li, K., Mutasa, S., Samothrakis, S., Wade, M., Walton, P.: AI Agents and Agentic Systems: A Multi-Expert Analysis. Journal of Computer Information Systems. 1–29 (2025). https://doi.org/10.1080/08874417.2025.2483832.
8. Woo, J.J., Yang, A.J., Olsen, R.J., Hasan, S.S., Nawabi, D.H., Nwachukwu, B.U., Williams, R.J., Ramkumar, P.N.: Custom Large Language Models Improve Accuracy: Comparing Retrieval Augmented Generation and Artificial Intelligence Agents to Noncustom Models for Evidence-Based Medicine. Arthroscopy: The Journal of Arthroscopic & Related Surgery. 41, 565-573.e6 (2025). https://doi.org/10.1016/j.arthro.2024.10.042.
9. Martins, A., Londral, A., L. Nunes, I., V. Lapão, L.: Unlocking human-like conversations: Scoping review of automation techniques for personalized healthcare interventions using conversational agents. International Journal of Medical Informatics. 185, 105385 (2024). https://doi.org/10.1016/j.ijmedinf.2024.105385.
10. Gaurav Samdani, Kabita Paul, Flavia Saldanha: Agentic AI in the Age of Hyper-Automation. World J. Adv. Eng. Technol. Sci. 8, 416–427 (2023). https://doi.org/10.30574/wjaets.2023.8.1.0042.
11. Sapkota, R., Roumeliotis, K.I., Karkee, M.: AI Agents vs. Agentic AI: A Conceptual Taxonomy, Applications and Challenges, https://arxiv.org/abs/2505.10468, (2025). https://doi.org/10.48550/ARXIV.2505.10468.
12. Ye, Y., Cong, X., Tian, S., Cao, J., Wang, H., Qin, Y., Lu, Y., Yu, H., Wang, H., Lin, Y., Liu, Z., Sun, M.: ProAgent: From Robotic Process Automation to Agentic Process Automation, https://arxiv.org/abs/2311.10751, (2023). https://doi.org/10.48550/ARXIV.2311.10751.
13. Lamanna, C.: Announcing new computer use in Microsoft Copilot Studio for UI automation, https://www.microsoft.com/en-us/microsoft-copilot/blog/copilot-studio/announcing-computer-use-microsoft-copilot-studio-ui-automation/.
14. Xiong: Invoice Extraction with OCR, https://github.com/MaxineXiong/Invoice-Extraction-OCR-Challenge-RPA.
15. RPA Challenge Invoice Extraction in UiPath | Automation OCR Challenge | UiPath RPA. (2020).
16. Willcocks, L.P., Lacity, M., Craig, A.: The IT function and robotic process automation. London School of Economics and Political Science, LSE Library (2015).
17. Syed, R., Wynn, M.T.: Robotic process automation: a review of the state-of-the-art. In: Grefen, P. and Vanderfeesten, I. (eds.) Handbook on Business Process Management and Digital Transformation. pp. 333–362. Edward Elgar Publishing (2024). https://doi.org/10.4337/9781802206098.00024.
18. Perković, G., Drobnjak, A., Botički, I.: Hallucinations in LLMs: Understanding and Addressing Challenges. In: 2024 47th MIPRO ICT and Electronics Convention (MIPRO). pp. 2084–2088. IEEE, Opatija, Croatia (2024). https://doi.org/10.1109/MIPRO60963.2024.10569238.

19. Chao, P., Robey, A., Dobriban, E., Hassani, H., Pappas, G.J., Wong, E.: Jailbreaking Black Box Large Language Models in Twenty Queries. In: 2025 IEEE Conference on Secure and Trustworthy Machine Learning (SaTML). pp. 23–42. IEEE, Copenhagen, Denmark (2025). https://doi.org/10.1109/SaTML64287.2025.00010.
20. Martino, A., Iannelli, M., Truong, C.: Knowledge Injection to Counter Large Language Model (LLM) Hallucination. In: Pesquita, C., Skaf-Molli, H., Efthymiou, V., Kirrane, S., Ngonga, A., Collarana, D., Cerqueira, R., Alam, M., Trojahn, C., and Hertling, S. (eds.) The Semantic Web: ESWC 2023 Satellite Events. pp. 182–185. Springer Nature Switzerland, Cham (2023). https://doi.org/10.1007/978-3-031-43458-7_34.
21. Spiess, C., Gros, D., Pai, K.S., Pradel, M., Rabin, M.R.I., Alipour, A., Jha, S., Devanbu, P., Ahmed, T.: Calibration and Correctness of Language Models for Code, https://arxiv.org/abs/2402.02047, (2024). https://doi.org/10.48550/ARXIV.2402.02047.
22. Schwartz, S., Yaeli, A., Shlomov, S.: Enhancing Trust in LLM-Based AI Automation Agents: New Considerations and Future Challenges, https://arxiv.org/abs/2308.05391, (2023). https://doi.org/10.48550/ARXIV.2308.05391.
23. van der Aalst, W.M.P., Bichler, M., Heinzl, A.: Robotic Process Automation. Bus Inf Syst Eng. 60, 269–272 (2018). https://doi.org/10.1007/s12599-018-0542-4.
24. Czarnecki, C., Fettke, P. eds: Robotic process automation: management, technology, applications. De Gruyter Oldenbourg, Berlin ; Boston (2021).

## Appendix 1

INVOICE DATA EXTRACTION TASK ENVIRONMENT:
- Web browser (Firefox) with invoice management systém
- Text editor (gedit) with example.csv file open

OBJECTIVE: Extract and compile invoice information into a structured CSV format.

DATA EXTRACTION PROCESS: For each invoice record:
INPUT DATA (from web interface):
- Invoice ID (unique identifier)
- Due Date (DD-MM-YYYY format)

EXTRACTION STEPS:
1. Click invoice download button for each ID
2. Open the downloaded invoice document
3. Extract the following fields:
   • Invoice Number
   • Invoice Date (DD-MM-YYYY format)
   • Company Name

- Total Amount Due

OUTPUT FORMAT (CSV):
ID,DueDate,InvoiceNo,InvoiceDate,CompanyName,TotalDue

SAMPLE DATA: 5jef1y8yx4t8yupbpo3fzg,25-02-2019,10021,13-02-2019,Sit Amet Corp.,1234.40     g11rzk7loegnitbclwnh8k,06-04-2019,223543,30-01-2019,Aenean LLC,976.00

Enter one complete record per line in the CSV file.